\documentclass[aip,apl
amsmath,amssymb,reprint]{revtex4-1}
\usepackage{graphicx}% Include figure files
\usepackage{caption}% left alignment of captions
\usepackage{dcolumn}% Align table columns on decimal point
\usepackage{bm}% bold math
\usepackage[utf8]{inputenc}
\usepackage[T1]{fontenc}
\usepackage{mathptmx}
\usepackage{etoolbox}
\usepackage{soul}%for st and ul command
\usepackage{subfig} % sub figures
\usepackage{multirow} %multi row in tables
\usepackage{pgfplotstable}
\usepackage{pgfplots}
\usetikzlibrary{matrix}%adjust horizontal spacing of legend in appendix plot

\makeatletter
\def\@email#1#2{%
 \endgroup
 \patchcmd{\titleblock@produce}
 {\frontmatter@RRAPformat}
 {\frontmatter@RRAPformat{\produce@RRAP{*#1\href{mailto:#2}{#2}}}\frontmatter@RRAPformat}
 {}{}
}%
\makeatother
\begin{document}

\preprint{AIP/123-QED}

\title[Identification and Mitigation of Conducting Package Losses for Quantum Superconducting Devices]{Identification and Mitigation of Conducting Package Losses for Quantum Superconducting Devices}
% Force line breaks with \\
\author{Yizhou Huang}
\affiliation{Department of Physics, University of Maryland, College Park, MD 20742, USA}
\affiliation{Maryland Quantum Materials Center, University of Maryland, College Park, MD 20742, USA}
\affiliation{Laboratory for Physical Sciences, 8050 Greenmead Drive, College Park, MD 20740, USA}

\author{Yi-Hsiang Huang}%
\affiliation{Laboratory for Physical Sciences, 8050 Greenmead Drive, College Park, MD 20740, USA}
\affiliation{Department of Electrical and Computer Engineering, University of Maryland, College Park, MD 20742, USA}

\author{Haozhi Wang}
\affiliation{Department of Physics, University of Maryland, College Park, MD 20742, USA}
\affiliation{Maryland Quantum Materials Center, University of Maryland, College Park, MD 20742, USA}
\affiliation{Laboratory for Physical Sciences, 8050 Greenmead Drive, College Park, MD 20740, USA}

\author{Zach Steffen}
\affiliation{Department of Physics, University of Maryland, College Park, MD 20742, USA}
\affiliation{Maryland Quantum Materials Center, University of Maryland, College Park, MD 20742, USA}
\affiliation{Laboratory for Physical Sciences, 8050 Greenmead Drive, College Park, MD 20740, USA}

\author{Jonathan Cripe}
\affiliation{Laboratory for Physical Sciences, 8050 Greenmead Drive, College Park, MD 20740, USA}

\author{F. C. Wellstood}
\affiliation{Department of Physics, University of Maryland, College Park, MD 20742, USA}
\affiliation{Maryland Quantum Materials Center, University of Maryland, College Park, MD 20742, USA}
\affiliation{Joint Quantum Institute, University of Maryland, College Park, MD 20742, USA}

\author{B. S. Palmer}
\email{bpalmer@umd.edu.}
\affiliation{Department of Physics, University of Maryland, College Park, MD 20742, USA}
\affiliation{Maryland Quantum Materials Center, University of Maryland, College Park, MD 20742, USA}
\affiliation{Laboratory for Physical Sciences, 8050 Greenmead Drive, College Park, MD 20740, USA}

\date{\today}% It is always \today, today,
 % but any date may be explicitly specified

\begin{abstract}
Low-loss superconducting rf devices are required when used for quantum computation. Here, we present a series of measurements and simulations showing that conducting losses in the packaging of our superconducting resonator devices affect the maximum achievable internal quality factors ($Q_i$) for a series of thin-film Al quarter-wave resonators with fundamental resonant frequencies varying between 4.9 and 5.8 GHz. By utilizing resonators with different widths and gaps, different volumes of the stored electromagnetic energy were sampled thus affecting $Q_i$. When the backside of the sapphire substrate of the resonator device is adhered to a Cu package with a conducting silver glue, a monotonic decrease in the maximum achievable $Q_i$ is found as the electromagnetic sampling volume is increased. This is a result of induced currents in large surface resistance regions and dissipation underneath the substrate. By placing a hole underneath the substrate and using superconducting material for the package, we decrease the ohmic losses and increase the maximum $Q_i$ for the larger size resonators. 
\end{abstract}

\maketitle

%\section{Introduction}
The ability to produce superconducting devices with low microwave loss and small phase noise is desired for both microwave kinetic inductance detectors and superconducting qubits. \cite{ZmuidzinasARCMP2012,deLeonScience2021} At the chip level, this requires the use of low-loss materials, clean fabrication processes, and good microwave hygiene.\cite{RichardsonSST2016,ucsb-airbridge,ibm-hidden-mode}
The packaging for the quantum chip should provide good impedance matching over a large bandwidth, a small amount of cross-talk between different signal lines, and good shielding to reduce radiated losses or prevent stray THz or IR black-body radiation from leaking into the package.\cite{mit-package-prx,CorcolesAPL2011} 

In this article, we measure limitations on the maximum achievable internal quality factors, $Q_{i}$, of a series of superconducting microwave resonators. The source of this loss is dissipation in normal metal conductors used in the package of the resonator chip. The energy stored in the resonator produces an rf magnetic field $H$ resulting in the production of shielding eddy currents in nearby conductors when $H$ impinges upon them. A noticeable amount of dissipation occurs when these shielding currents are produced in conductors with a finite surface resistance. This loss mechanism was initially identified when measuring five resonators on a single chip. A systematic decrease in their quality factors was observed to correlate with an increase in the widths and gaps of the resonators. To model this finding, we performed finite-element simulations to estimate the magnitude of dissipation in each conductor used in the package. A conducting adhesive, used to adhere and thermalize the chip to the package, was identified to be the most significant source of loss. By implementing a few changes to the packaging, we reduced this loss and demonstrated over an order of magnitude improvement in the maximum $Q_i$ of the resonators.

%\section{Device and Results}
The thin film Al chip that we measured consists of five multiplexed quarter-wave coplanar waveguide (CPW) resonators coupled to a common coplanar waveguide transmission feedline.~\cite{GaoAPL2008} The resonators had different fundamental resonant frequencies $f_{\circ}$, CPW widths $w$ and gaps $g$ ranging from $f_{\circ}=$ 4.9 \text{GHz}, $w=3\ \mathrm{\mu m}$ and $g = 1.5\ \mathrm{\mu m}$ for R1 up to $f_{\circ}=$ 5.8 \text{GHz}, $w=22\ \mathrm{\mu m}$ and $g = 11\ \mathrm{\mu m}$ for R5 (see supplemental material for more details).

\begin{figure}
\captionsetup{singlelinecheck = false, justification=centerlast}
\subfloat[ ]
{
\centering
\includegraphics[width=0.45\linewidth]{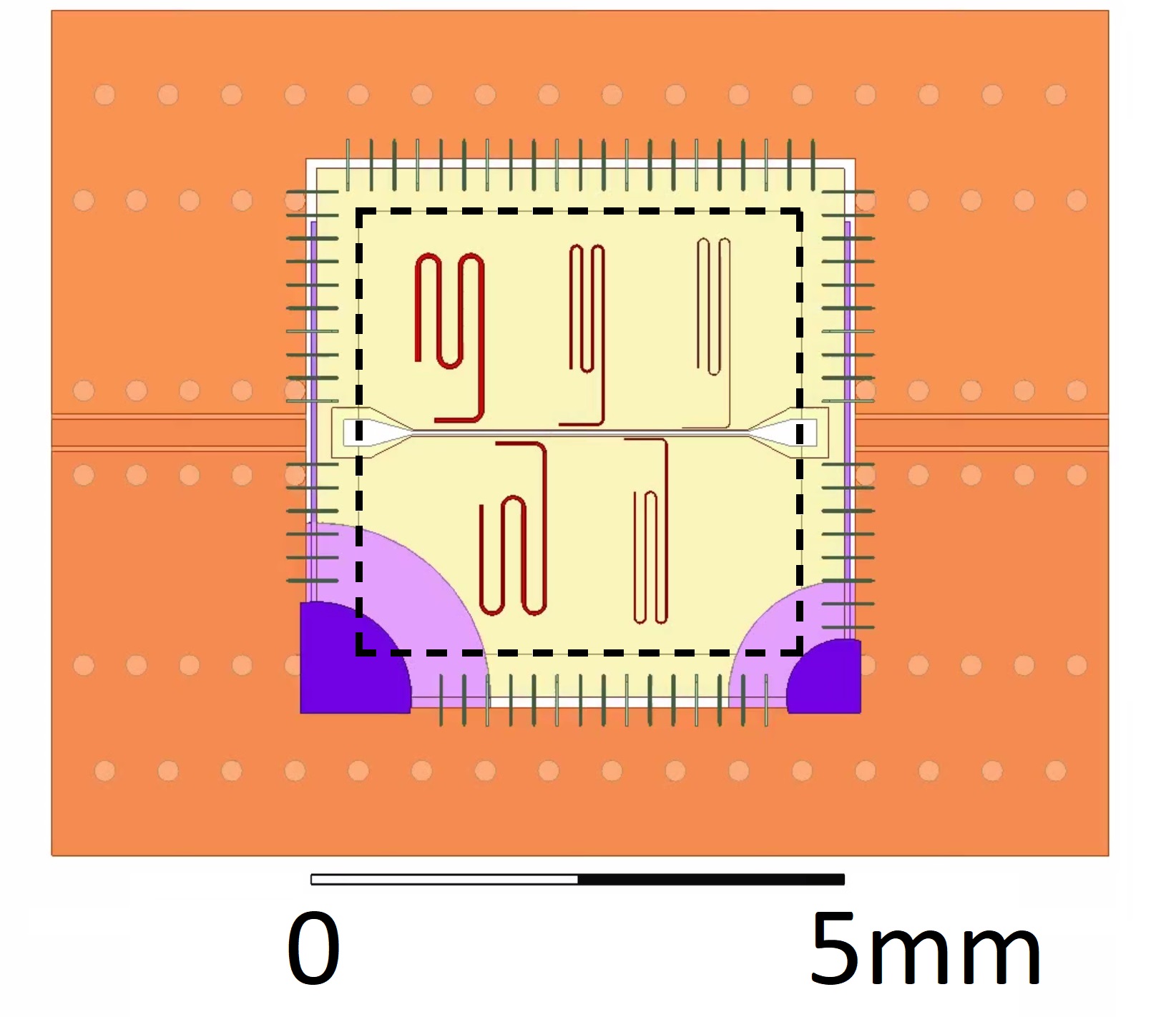}
}
\hfill
\subfloat[ ]
{
\centering
\includegraphics[width=0.45\linewidth]{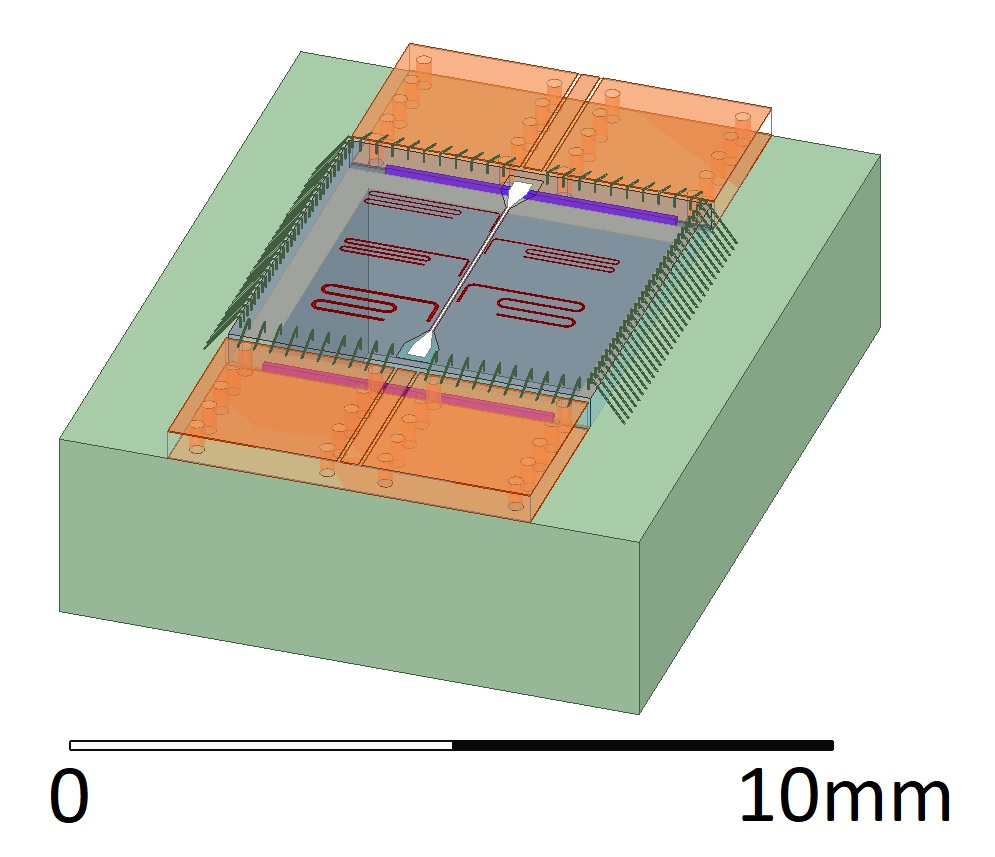}
}

\caption{\label{fig:hfss_setup}CAD rendering of resonator chip and surrounding PCB (lid and sidewall not shown). (a) Plan view of device in Cu packages, showing substrate (light yellow), resonators (red), center transmission line (white), surrounding PCB (orange). A series of connections was used to represent the wirebond connections from the chip to the surrounding PCB. For the Cu$\blacksquare$ package, glue was present underneath the area of the chip. For the Cu$\square$ package, areas of glue above (dark purple) and below (light purple) the substrate are shown here as well as the size and approximate location of the hole (dashed contour). (b) Trimetric view of chip and Al$\square$ package.}
\end{figure}

The same chip was sequentially packaged in four different ways and measured. To efficiently conduct heat from the chip, a silver impregnated conducting adhesive,\footnote{Loctite Ablestik: 59C} diluted with toluene, was used to attach the chip in each package. For the first measurement, the backside of the chip was glued to an oxygen-free high thermal conductivity (OFHC) Cu package (denoted Cu$\blacksquare$). A two layer Cu printed circuit board (PCB), which was soldered to the OFHC Cu package, was used to interface the rf signals from a non-magnetic SMA connector to the resonator chip (see Fig.\ref{fig:hfss_setup} (a) for a representative CAD rendering).

To measure the low-temperature loss of the resonators, the packaged device was bolted to the mixing chamber stage of a Leiden cryogen-free dilution refrigerator and connected to an input and output microwave cable (See Ref.\onlinecite{yeh-fridge-setup} for further details of the set-up). To reduce stray magnetic fields, the device was located near the bottom of two open ended magnetic shield cylinders.\footnote{Amuneal, Amumetal 4K (A4K). The cylinders are about 40cm long.} All of the $Q_{i}$ data presented here was with the refrigerator at its base temperature and the mixing chamber less than 20 mK. A vector network analyzer was used to measure the in-phase and out-of-phase ratio of the transmitted voltage to input voltage at 1601 different discrete frequencies ($S_{21}(f)$) spanning the resonance and at different input drive voltages. Each $S_{21}(f)$ scan was repeated multiple times, from which the mean $\bar{S}_{21}(f)$ and the standard deviation $\sigma_{S_{21}}(f)$ at each frequency were calculated for both quadratures. Both quadratures of the mean $\bar{S}_{21}(f)$ were simultaneously fitted, weighted by $1/\sigma_{S_{21}}(f)$, using the diameter-correction method\cite{dcm} to extract 5 fitting parameters including $Q_i$. 

\begin{figure}[htbp]
{\centering
\includegraphics[page=1]{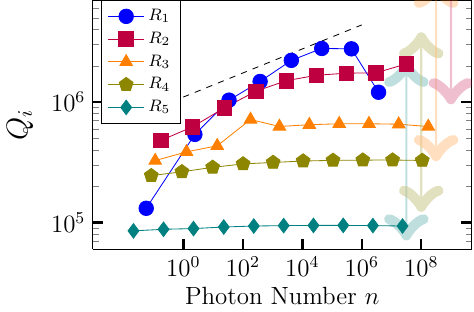}\par}
\caption{Log-log plot of the fitted resonator internal quality factors versus stored photon number for $\text{Cu}\blacksquare$ package. 
%R1 and R2 shows a weak power dependence. 
As the width and gap of the resonator increases from R1 to R5, the measured power dependence is smaller and the maximum $Q_{i,m}$ decreases. The dashed line on top is a weak $\propto n^{0.1}$ power law as a guide. The semi-transparent arrows are bounds of the estimated $Q_{i,m}$ for $R_2$ to $R_5$, using Eq.\ref{eq:qi_h}, and the geometric factors $\gamma$ from the first column of Table \ref{tab:hfss_ratio}. The lower bound assumes $\rho_{glue}$ for the resistivity underneath the substrate while the upper bound assumes $\rho_{Cu}$ for the upper bound.}\label{fig:cu_nohole_all5}
\end{figure}

Fig. \ref{fig:cu_nohole_all5} shows a log-log plot of the fitted $Q_i$ versus stored average photon number from the first measurement of the resonators in Cu$\blacksquare$ package. For R1, a weak increase in $Q_i$ with increasing power was observed with a maximum $Q_{i,m} \simeq 2 \times 10^6$. As $w$ and $g$ of the resonators increase, the observed power dependence decreases and $Q_{i,m}$ decreases to $Q_{i,m} = 10^5$ for resonator R5. 

The focus of this paper is to determine the physical mechanism responsible for the limitations on $Q_{i,m}$. These limitations are not consistent with losses at the interfaces near the resonator because $Q_{i,m}$ decreases with an increase in $g$ and $w$.\cite{surface-woods} Instead, loss farther from the resonator was thought to be the source. Since normal metal conductors were underneath and surrounded the resonator chip, our conjecture was that Ohmic dissipation from these normal metal components limited $Q_{i,m}$.

%--------- BEGIN FIGURE--------------

%------------- END FIGURE --------------------

\begin{figure}[htbp]
{\centering
\includegraphics[page=1]{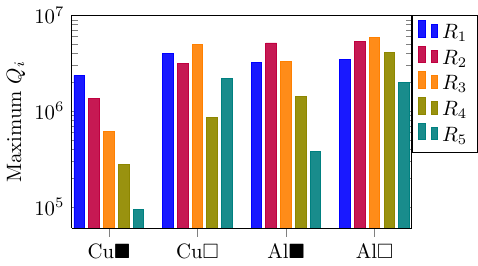}\par}
\caption{Bar chart of measured maximum $Q_{i,m}$ on a log scale for each resonator in its corresponding Cu or Al package, without a hole ($\blacksquare$) and with a 4.2 x 4.2 mm$^2$ hole ($\square$) underneath the 5 x 5 mm$^2$ chip.
}\label{fig:max_qi_all5}
\end{figure}

To examine this hypothesis, we modified the package and remeasured the same resonator chip. First, a 4.2 mm $\times$ 4.2 mm wide hole that was 2.5 mm deep was milled out of the bottom of the Cu package where the chip resided.\cite{lienhard2019microwave} To adhere the device, glue was added to two corners of the chip and along the sides (see purple regions in Fig.~\ref{fig:hfss_setup}(a)). With these modifications (denoted $\text{Cu}\square$), all of the $Q_{i,m}$'s improved, especially for R5 which exhibited over a factor of 20 improvement (see Fig. \ref{fig:max_qi_all5}). In addition, there was no longer a strong correlation of $Q_{i,m}$ to $w$ and $g$ of the resonators.

%------------- BEGIN FIGURE --------------------

%------------- END FIGURE --------------------

To determine whether the $Q_{i,m}$'s could be improved further, two packages with overall the same geometry were manufactured from aluminum 6063 and used to measure the same resonator chip. One of them had no hole (Al$\blacksquare$) and one had a hole (Al$\square$). Similar to the Cu$\square$ case, a smaller amount of glue was applied underneath or on the perimeter of the chip for both Al packages. Also, the Cu PCB, which was glued with a silver epoxy\footnote{CircuitWorks: CW2400} to the Al package, was trimmed and cut into two pieces so that it was only within close proximity to the side of the chip with the input and output SMA launchers (see Fig.\ref{fig:hfss_setup}(b) and supplemental material). While the $Q_{i,m}$'s for the Al$\blacksquare$ showed an overall increase over the $Q_{i,m}$'s measured in Cu$\blacksquare$, an increase in the loss with resonator size from R2 to R5 was still observed. The data in the Al$\square$ package had similar $Q_{i,m}$'s as measured for Cu$\square$. Next, we discuss a model and the use of microwave finite element simulations to identify the observed sources of loss in our package. 

%--------------- MODEL AND SIMULATIONS---------------------

%\section{Model \& Simulations}%\label{sec:sim}
Dissipation from induced eddy currents in the surrounding normal metal associated with the packaging was hypothesized as the source limiting $Q_{i,m}$. Neglecting the anomalous skin effect, ac shielding currents in a normal metal with resistivity $\rho$ decay on a length scale given by the skin depth $\delta = \sqrt{\frac{2\rho}{2\pi f \mu_{\circ}}}$, this results in an effective surface resistance $R_{S} = \rho/\delta$.\cite{Padamsee,Pozar} The power dissipated in $R_S$ also produces a limitation in the internal quality factor given by \cite{Padamsee}
\begin{equation}
 Q_{i,R_S}^{-1}= \frac{R_{S}}{2\pi f_{\circ}\mu_{\circ}} \gamma= \frac{R_{S}}{2 \pi f_{\circ}\mu_{\circ}}\frac{\int\int_{S}|H|^{2}dS}{\int\int\int_{V}|H|^{2}dV}.\label{eq:qi_h}
\end{equation}
Here, the ratio of the two integrals, which we define as $\gamma$, is a geometric factor and equals the ratio of the magnetic field energy at the surface of $R_{S}$ to the total magnetic field energy. For scaling purposes, $f_{\circ}= 5.6$ GHz for R4 and the ratio $R_{S}/(2\pi f_{\circ}\mu_{\circ}) = (3.4\times10^{-3}\sqrt{\rho}) \ \mathrm{[m/\Omega]}^{1/2}$. 

There are three normal metal conductors of concern present in our packaging. First, the Ag impregnated glue was measured to have a relatively large dc resistivity of $\rho_{\text{glue}} = 630\ \mu\Omega\cdot$cm at $T = 77$ K. For the Ag glue, the ratio $R_{S,glue}/(\omega_{\circ}\mu_{\circ})= 8.5 \times 10^{-6}\ \mathrm{m}$ at 5.6 GHz, implying $\gamma_{glue} < (1/8.5) \ \mathrm{m}^{-1}$ to achieve $Q_{i} > 10^{6}$. The other normal conductor of concern was the Cu used in the two Cu packages. From quality factor measurements of a 3D OFHC Cu cavity at $ T=3$ K, the resistivity of the Cu is estimated to be $\rho_{\text{Cu}}\simeq 0.6\ \mu\Omega\cdot$cm. The third and final conductor was the PCB Cu. By manufacturing a bandpass microwave CPW resonator from a similar PCB and measuring $Q$ at $T = 3$ K, a dc resistivity $\rho_{\text{PCB}} \simeq 2\ \mu\Omega\cdot$cm was estimated.\cite{guptamicrostrip} These three values of $\rho$ are used to estimate the loss.

To calculate the geometric factor $\gamma$ at the surface of each conductor of concern, the $H$ field for each resonator in the different package geometries was simulated using Ansys' high frequency simulation software (HFSS). Each conductor in the simulation was assumed to be a perfect electric conductor to reduce simulation resources. To simulate the $H$ field associated with the stored energy of the resonator, and not with the coupling to the CPW transmission line, we implemented two effects in the simulation. First, the connections between the signal trace of the resonator chip and the PCB were removed. This prevented radiation from leaking to the ports where the SMA connectors were located and also avoided a standing wave near a resonant frequency. Second, the open end of the quarter-wave resonator was shunted with an excitation lumped port with a matched impedance to the resonator waveguide, which is 50 $\Omega$. The resonator was then excited at $f_{\circ}$, which was determined by satisfying $\mathrm{Im}\left[(Y_{11}(f_\circ)\right] = 0$ where $Y_{11}$ is the self-admittance of the lumped port. We simulated each resonator in the four different packages and calculated $\gamma$ for each conductor of concern.

\begin{table}[htbp]
\captionsetup{singlelinecheck = false, justification=centerlast}
\caption{HFSS simulated magnetic field geometric factors $\gamma$ in units of $\mathrm{m}^{-1}$, for the different size resonators and in the various packages. Three surfaces were considered for the geometric factor calculations: the Cu packaging ``Base'', the ``PCB'' Cu, and regions where ``Glue'' was used to adhere the device. $\gamma$ factors smaller than $0.005\ \mathrm{m}^{-1}$ are left blank.}\label{tab:hfss_ratio}
\begin{tabular}{|c|ll|lll|ll|ll|}
\hline
\multirow{2}{*}{R\#} & \multicolumn{2}{c|}{Cu$\blacksquare$} & \multicolumn{3}{c|}{Cu$\square$} & \multicolumn{2}{c|}{Al$\blacksquare$} & \multicolumn{2}{c|}{Al$\square$} \\ \cline{2-10} 
 & \multicolumn{1}{c|}{$\gamma_{\text{Base/Glue}}$} & $\gamma_{\text{PCB}}$ & \multicolumn{1}{c|}{$\gamma_{\text{Base}}$} & \multicolumn{1}{l|}{$\gamma_{\text{PCB}}$} & $\gamma_{\text{Glue}}$ & \multicolumn{1}{c|}{$\gamma_{\text{PCB}}$} & $\gamma_{\text{Glue}}$ & \multicolumn{1}{c|}{$\gamma_{\text{PCB}}$} & $\gamma_{\text{Glue}}$ \\ \hline
R1 & \multicolumn{1}{c|}{0.02} & 0.01 & \multicolumn{1}{c|}{-} & \multicolumn{1}{c|}{-} & \multicolumn{1}{c|}{-} & \multicolumn{1}{c|}{-} & \multicolumn{1}{c|}{-} & \multicolumn{1}{c|}{-} & \multicolumn{1}{c|}{-} \\ \hline
R2 & \multicolumn{1}{c|}{0.12} & 0.06 & \multicolumn{1}{c|}{0.01} & \multicolumn{1}{c|}{-} & \multicolumn{1}{c|}{-} & \multicolumn{1}{c|}{0.01} & \multicolumn{1}{c|}{-} & \multicolumn{1}{c|}{-} & \multicolumn{1}{c|}{-} \\ \hline
R3 & \multicolumn{1}{c|}{0.35} & 0.16 & \multicolumn{1}{c|}{0.02} & \multicolumn{1}{c|}{0.02} & \multicolumn{1}{c|}{-} & \multicolumn{1}{c|}{0.06} & 0.02 & \multicolumn{1}{c|}{0.01} & \multicolumn{1}{c|}{-} \\ \hline
R4 & \multicolumn{1}{c|}{0.95} & 0.42 & \multicolumn{1}{c|}{0.06} & \multicolumn{1}{c|}{0.04} & 0.12 & \multicolumn{1}{c|}{0.24} & 0.07 & \multicolumn{1}{c|}{0.03} & 0.01 \\ \hline
R5 & \multicolumn{1}{c|}{1.6} & 0.89 & \multicolumn{1}{c|}{0.16} & \multicolumn{1}{c|}{0.13} & 0.04 & \multicolumn{1}{c|}{0.77} & 0.22 & \multicolumn{1}{c|}{0.12} & 0.03 \\ \hline
\end{tabular}
%\par}
\end{table}

Table \ref{tab:hfss_ratio} presents the calculated geometric factors $\gamma$ associated with the different normal conducting regions for each resonator in the different packages. These regions were the OFHC Cu material (``Base''), the Ag impregnated glue, and the PCB Cu used to feed signals to the resonator device. Because glue was present underneath the entire substrate of the chip in $\text{Cu}\blacksquare$ package, we have combined base and glue together for that measurement. For other packages, pictures of the device were used to identify the area covered by the glue to calculate $\gamma$ (\emph{e.g.}, see purple regions in Fig.~\ref{fig:hfss_setup}). 

\begin{figure}[htbp]
	\captionsetup{singlelinecheck = false, justification=centerlast}
	\includegraphics[width=0.95\linewidth]{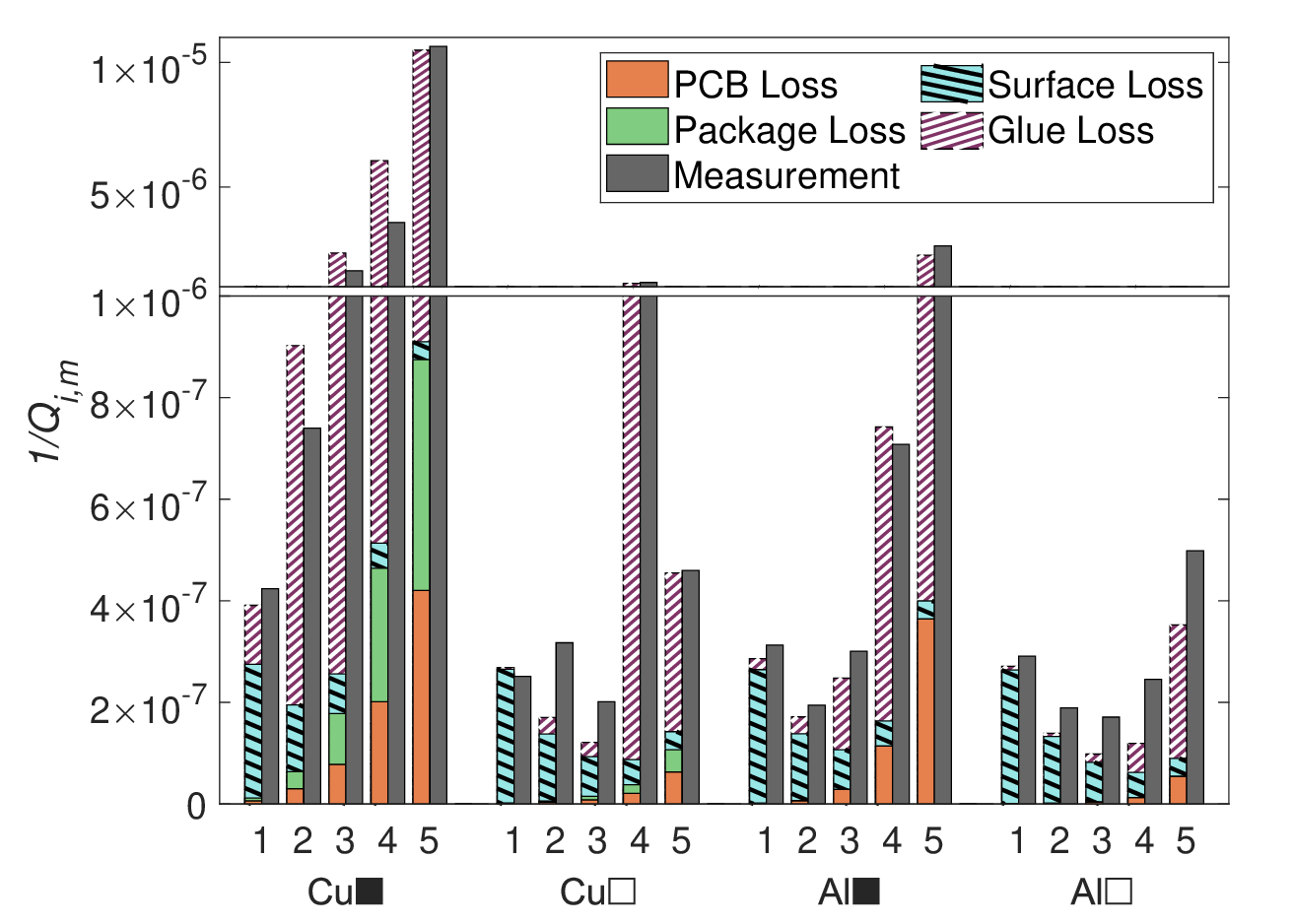}
	\caption{\label{fig:simulate_and_measurement}Comparison of estimated and measured resonator loss $1 / Q_{i,m}$ for each resonator and in each package. Note the scale of the y-axis changes at $1 \times 10^{-6}$.}
\end{figure}

Using the corresponding geometric values from Table \ref{tab:hfss_ratio} and the measured resistivity values, individual losses due to the glue, Cu package, and PCB were calculated using Eq.~\ref{eq:qi_h} and compared against measured values in the bar chart of Fig.\ref{fig:simulate_and_measurement}. Note that for $\text{Cu}\blacksquare$ package, a 70\% fill rate for the glue was assumed. For R5, for example, in Cu$\blacksquare$, $0.7\times\gamma_{glue} = 1.2\ \mathrm{m}^{-1}$ implying $Q_{i,R_{S}} \simeq \times 10^{5}$, a value comparable to the observed measured loss. Loss from the glue explains the $w$ and $g$ trend observed in the Cu$\blacksquare$ package and the apparent random high loss measured with R4 and R5 in the other three packages. We note that Goetz \textit{et al.} reached a similar conclusion for their source of loss; namely that it was associated with a conducting adhesive.\cite{goetz2016loss} 

Despite these losses, several benefits of the hole underneath the chip are noted: 
\begin{enumerate}
 \item The induced currents in the PCB are reduced due to the stored $H$ field residing in vacuum underneath the device (\emph{e.g.}, compare $\gamma_{\text{PCB}}$ for Al$\blacksquare$ and Al$\square$).
 \item The amount of rf current flowing on and off the chip through the wire bonds is decreased.
 \item In simulation, the fundamental resonant frequency of an undesired package EM mode, which can couple to qubits and result in decoherence,\cite{ibm-hidden-mode} was shown to increase from approximately 11 GHz up to 22 GHz due to the lower dielectric constant of the vacuum in the hole, thus resulting in a smaller interaction with the qubit. 
\end{enumerate}

Finally, as shown in Fig.~\ref{fig:simulate_and_measurement}, a slight increase in $Q_{i,m}$ with an increase in resonator width and gap for R1 to R3 in Al$\square$ was observed. To account for this small trend, a thin layer of dielectric loss with loss tangent $\delta$ at the interfaces was simulated using COMSOL. For this simulation of ``surface loss'', layers with thickness $t$ and relative dielectric constant $\epsilon_r$ were fixed such that $\left(t\times \delta / \epsilon_r \right) \sim 10^{-5}$ nm (see supplementary materials for details).\cite{surface-woods}

%\section{Conclusion}
In conclusion, conductive losses associated with resistive materials used in the packaging of the device resulted in an increase in the internal losses of superconducting microwave resonators. To explore the source of these losses, a resonator chip was measured in four different packages and finite-element microwave simulations were performed. 
Our measurements and simulations show that resonators with wider center line traces and gaps induced a larger amount of eddy shielding currents in conducting material directly below the substrate of the chip and that this could limit the resonator's $Q_i$ when that conductor had a large resistivity. A predominant source of loss was the silver impregnated glue that was used as the adhesive between the substrate and the package base. This loss and loss from a normal metal conducting PCB surrounding the device can be mitigated by creating a hole in the package directly below the chip and using material with a smaller surface resistance. 

While the effect of different packages on the quality factors of superconducting coplanar waveguide resonators was measured and simulated in this paper, our simulations can be extended to superconducting transmon qubits. In particular, an x-mon qubit \cite{ucsb-xmon} with $w = g = 30 \ \mathrm{\mu m}$ and a fundamental resonance at $f_{\circ} = 6\ \text{GHz}$ was simulated. The center of the x-mon was placed $1.25\ \text{mm}$ away from both edges of the substrate in $\text{Cu}\blacksquare$, and in the absence of the conducting glue a $T_1\ \mathrm{of}\ 3\ \mathrm{\mu s}$ limited by the surface resistance of the OFHC Cu backing was found. Switching to $\text{Al}\square$ would increase $T_1\ \mathrm{to}\ 80\ \mathrm{\mu s}$ limited by the Cu in the PCB. Furthermore, to downconvert hot phonons and reduce quasiparticle tunneling charge parity rates, a few groups have electroplated Cu on the backside of the substrate in a grid of squares.\cite{mcdermott-phonon-qubit} For a $0.5 \times 0.5$\ mm$^{2}$ Cu grid with a 50\% fill rate, enough induced currents are produced such that $T_{1}\ \mathrm{was}\ 18\ \mu \text{s}$ for the simulated x-mon considered here.

Based on the results of this paper, a few final recommendations are made:
\begin{enumerate}
\item The use of normal conducting glues is not recommended; better choices are the use of dielectric glues or no glue.\footnote{Qdevil: Qcage} 
\item Increasing the separation between the resonator or qubit device and normal conducting material with the use of holes underneath the substrate or the use of thicker substrates significantly reduces currents underneath as well as on and off the device through the wirebonds.
\item The use of superconductors with a smaller surface resistance in the package and PCB could be essential to decrease the loss in very low loss future qubits or resonators. 
\item Simulations similar to the ones performed here to estimate $\gamma$ and knowledge of the microwave surface resistance of the materials used in the package are informative . 
\end{enumerate}

\section*{Supplementary Material}

Additional materials, such as details of fabrication steps, pictures of devices, samples of fit for $S_{21}(f)$ and fitted $Q_i$ versus photon number in packages other than $\text{Cu}\blacksquare$ are contained in the supplementary materials.

\begin{acknowledgments}
The authors acknowledge useful suggestions and conversations with Christopher Lobb. The authors thank Danielle Braje and MIT Lincoln Laboratory for the design of the resonator chip and Ashish Alexander and Chris Richardson for the use of their PCB.
\end{acknowledgments}

\section*{Data Availability Statement}
The data that support the findings of this study are available from the corresponding author upon reasonable request.

\nocite{*}
\bibliography{aip-revised}% Produces the bibliography via BibTeX.

\end{document}

\title{Supplementary Materials for Identification and Mitigation of Conducting Package Losses for Quantum Superconducting Devices}

\maketitle

\section{Device Fabrication}

Our device was fabricated from a 3-inch diameter Kyocera sapphire wafer with a thickness of 0.43mm. After the wafer was cleaned in a 3:1 solution of $\text{H}_2\text{SO}_4$:$\text{H}_2\text{O}_2$, rinsed with water, IPA and dried, the wafer was loaded into a Plassys 550 MEB evaporator. After reaching a base pressure $\sim 10^{-8}\ \text{mBar}$, the wafer holder was heated to 700\textdegree{} C for 30 minutes before cooling down to 200\textdegree{} C. 
A Ti getter was evaporated in the chamber, before depositing a 100 nm thick  Al layer at an approximate rate of 0.4 nm/s with a background pressure of approximately $2\times 10^{-7}$ mBar. 

To define the resonator pattern, Fujifilm 906-10 positive photoresist was spun on the wafer, pre-baked, exposed with an i-line stepper, and developed with OPD4262. The aluminum was then etched with commercial aluminum etchant (80\% Phosphoric Acid, 10\% Acetic Acid, 2\% Nitric Acid, \& 8\% H$_2$O, J.T. Baker) at room temperature. The remaining 906-10 photoresist was removed in two baths of Microdeposit Remover 1165 at 80\textdegree{} C for 60 and 30 minutes each. A protective FSC-M resist was spun onto the wafer before it was diced into  $5\ \text{mm} \times 5\ \text{mm}$ dies. The FSC-M resist was removed in acetone, methanol and IPA before the chip was mounted into the first Cu$\blacksquare$ package.

%The device was fabricated by evaporating a 100 nm thick film of Al on a pirana cleaned and in-situ vacuum annealed 0.43 mm thick sapphire substrate. After deposition of the Al film, photoresist was spun, pre-baked, exposed using a stepper, post-baked and then developed. The Al film was then etched using aluminum etchant (80\% Phosphoric Acid, 10\% Acetic Acid, 2\% Nitric Acid, \& 8\% H$_2$O), the photoresist was removed and a protective resist was applied before the wafer was diced into the  $5 \text{mm} \times 5 \text{mm}$ device die. 

\begin{figure}
\subfloat[]
{%\centering
\includegraphics[width=0.357\linewidth]{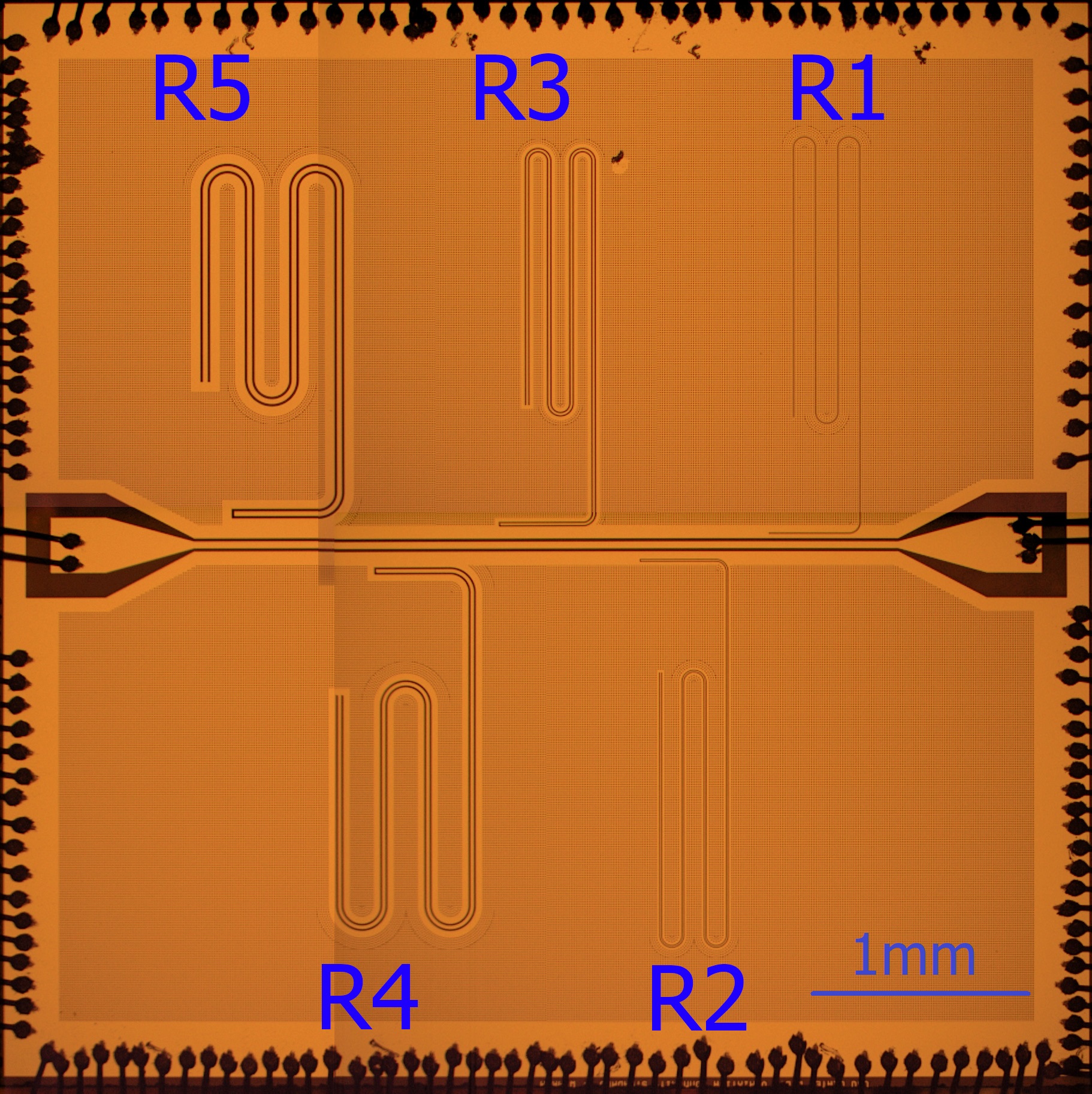}\label{fig:device_micrograph}
}
\hfill
\subfloat[]
{%\centering
\includegraphics[width=0.2834\linewidth]{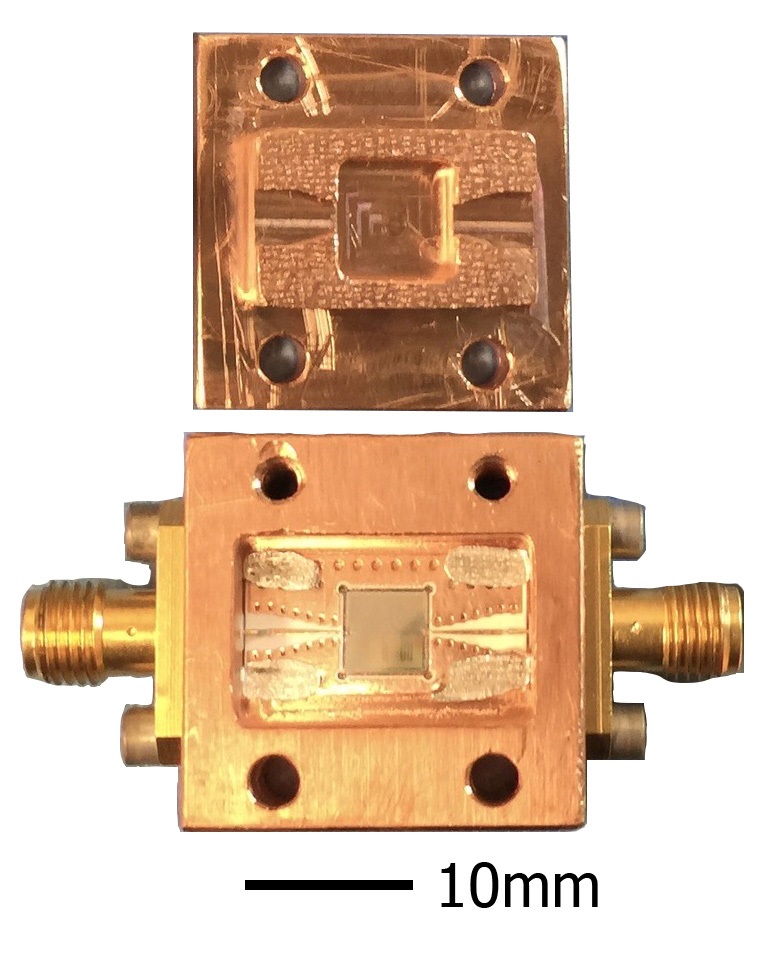}\label{fig:device_photo_cu}
}
\hfill
\subfloat[]
{%\centering
\includegraphics[width=0.309\linewidth]{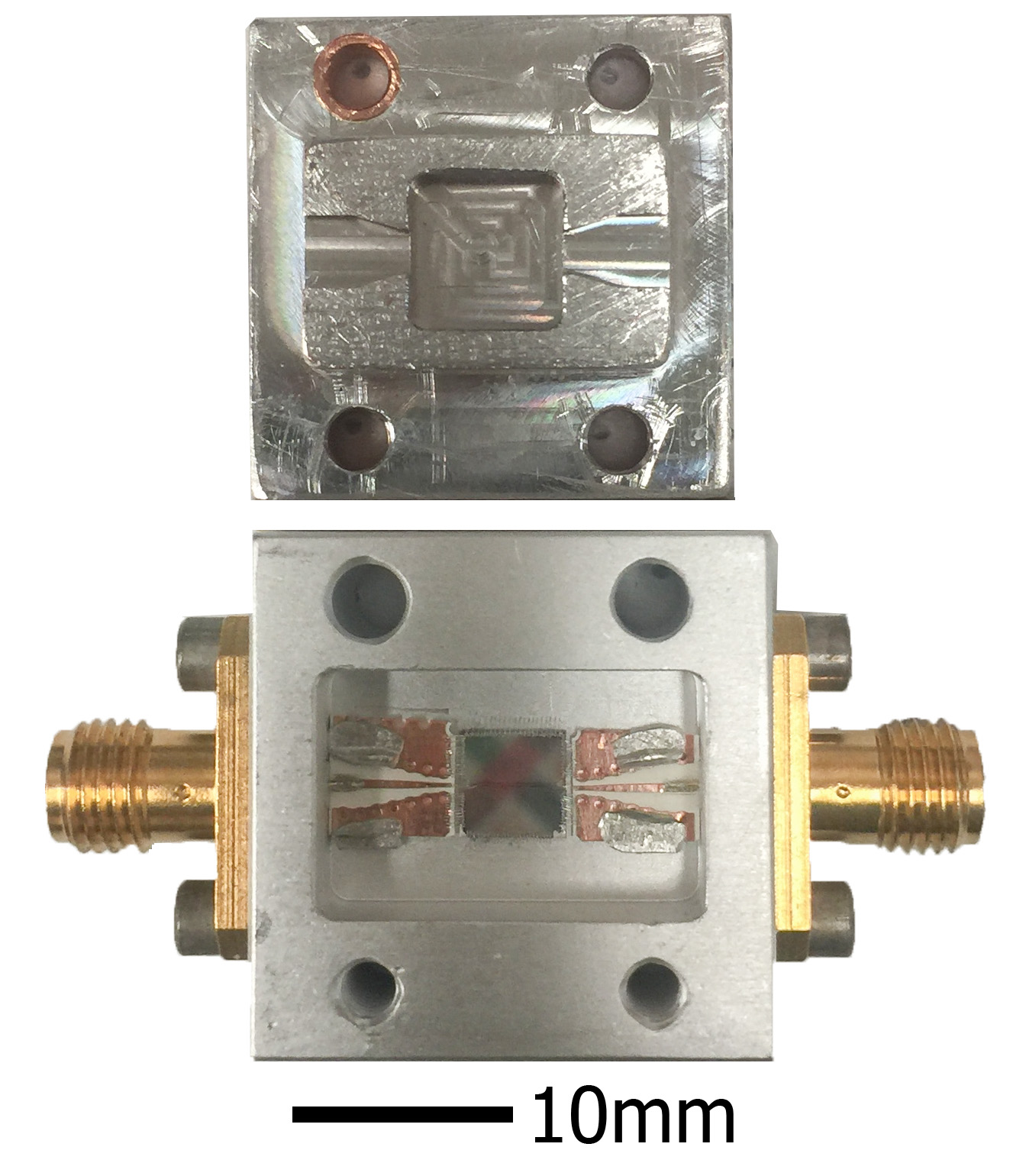}\label{fig:device_photo_al}
}
\caption{Photos of the chip and two of the packaged devices. (a) Stitched micrograph %Confocal laser image 
of the device with  resonator denotations added. (b) Picture of device base (bottom) and lid (top) in a Cu package with no hole ($\text{Cu}\blacksquare$) and (c) in an aluminum package with no hole ($\text{Al}\blacksquare$). The chip was wirebonded to a PCB with Si doped Al wires, which were in turn soldered to non-magnetic SMA connectors. A small length of indium wire was placed on the ground of the PCB before bolting the lid of the package box in place.
}
\label{fig:device_photo}
\end{figure}

A list of the design parameters for all five resonators is shown in Table \ref{tab:design}. Fig.\ref{fig:device_photo} shows a micrograph of the chip (a), the chip in the Cu$\blacksquare$ package (b), and the chip in the Al$\blacksquare$ (c).

\begin{table}[htbp]
{\centering
\caption{Resonator Design Parameters}\label{tab:design}
\begin{tabular}{|r|c|c|c|c|}
\hline
R\# & w ($\mu m$) & g ($\mu m$) & $f_\circ$ (GHz) & $Q_c$ \\ \hline
1   & 3            & 1.5       & 4.9    &   250k     \\ \hline
2   & 6            & 3         & 5.2    &   290k     \\ \hline
3   & 10           & 5         & 5.4    &   310k     \\ \hline
4   & 16           & 8         & 5.6    &   260k     \\ \hline
5   & 22           & 11        & 5.8    &   210k     \\ \hline
\end{tabular}\par}
\end{table}

\section{Measurement Data}

\begin{figure}[htbp]
	\centering
	\includegraphics[width=0.8\linewidth]{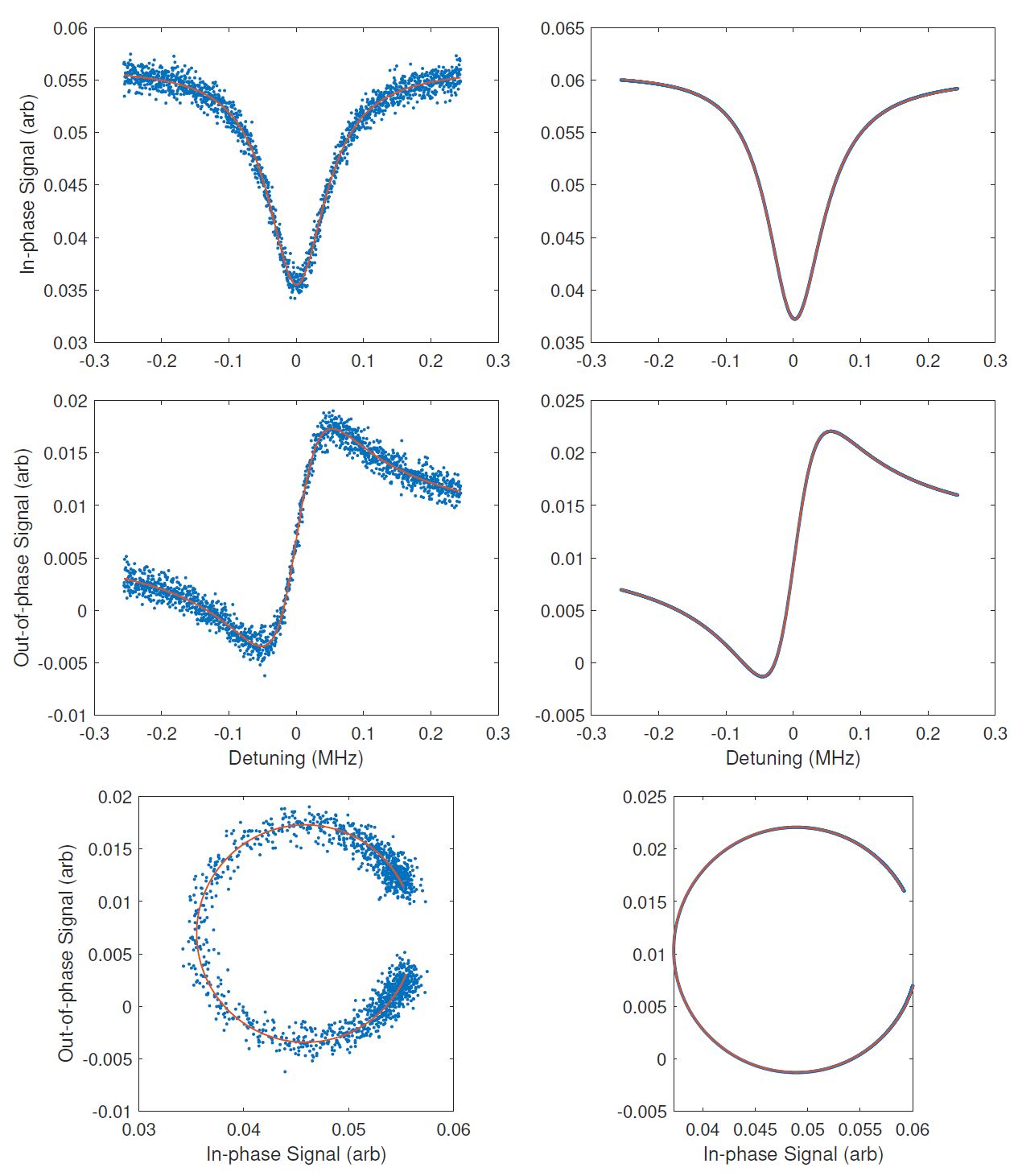}
	\caption{Sample data ($\bar{S}_{21}(f)$, blue points) and fit (red line) using the diameter-correction method after correcting for electrical delay, for R5 in $\text{Cu}\blacksquare$ package. The three left panels are measured at -70dBm probe power (corresponding to $\langle n \rangle \sim 0.2$ resonator photon), the three right panels are measured at 10dBm probe power (corresponding to $\langle n \rangle \sim 2\times 10^7$ resonator photon). The top two panels are the in-phase part of $\bar{S}_{21}(f)$ versus frequency, plotted as detuning from resonant frequency $f_\circ \simeq 5.835796\text{GHz}$, the middle two panels are out-of-phase part of $\bar{S}_{21}(f)$ versus frequency, and the bottom two panels are in-phase part of $\bar{S}_{21}(f)$ versus out-of-phase part. The reduced $\chi^2$ is $1.008$ for the panels on the left, and $\sim 46$ for the panels on the right.}\label{fig:s21_sample}
\end{figure}

The device in the main text was measured over a 6 month duration in the following order of four packages: $\text{Cu}\blacksquare, \text{Cu}\square,\  \text{Al}\square,\ \text{and}\ \text{Al}\blacksquare$.
A Keysight E5071C vector network analyzer (VNA) was used to measure $S_{21}(f)$ of the resonators.  The intermediate frequency (IF) bandwidth and the number of measurement repetitions at each probe power were set to balance data acquisition speed and signal-to-noise ratio, using the measurement parameters provided in Table \ref{tab:vna_setting}. To illustrate the quality of the fit, in Fig.~\ref{fig:s21_sample} we present two sets of data $\bar{S}_{21}(f)$ and their fits for R5 in $\text{Cu}\blacksquare$ package, one at low photon number and the other at high photon number.

\begin{figure}
{\centering
\begin{tikzpicture}[]
\begin{axis}[
	width = 0.7\linewidth, height = 0.5\linewidth,
	%xlabel=\large Photon Number $n$,
    xmode=log,
        xmax = 1*10^9,xmin = 3*10^-2,xtick={1,10^2,10^4,10^6,10^8},ymax = 1.7*10^7,ymin = 1*10^5,
	ylabel=$Q_i$,ymode = log,
	legend style={at={(0.97,0.97)},anchor=north east, /tikz/every even column/.append style={column sep=0.023\linewidth}},legend columns=-1,
	axis x line = bottom,
	minor x tick style={draw=none},
	mark options={mark size=3.5pt},% Ben demands more distinguishable markers
    xticklabels={,,,,},% hide x tick label for this figure
	y tick label style={font=\large}, label style={font=\Large},% Ben demands larger size
        major tick style={very thick, major tick length=5pt, black}, % Ben demands bolder and more visible tickets
        tick align=inside,
	execute at begin axis={\draw (rel axis cs:0,1) -- (rel axis cs:1,1);},
]
\addplot[color = blue, mark = *] table [x=<n>,y=Q_i]{CuHole_R1.txt};
\addlegendentry{$R_1$};
\addplot[color = purple, mark = square*] table [x=<n>,y=Q_i]{CuHole_R2.txt};
\addlegendentry{$R_2$};
\addplot[color = orange, mark = triangle*] table [x=<n>,y=Q_i]{CuHole_R3.txt};
\addlegendentry{$R_3$};
\addplot[color = olive, mark = pentagon*] table [x=<n>,y=Q_i]{CuHole_R4.txt};
\addlegendentry{$R_4$};
\addplot[color = teal, mark = diamond*] table [x=<n>,y=Q_i]{CuHole_R5.txt};
\addlegendentry{$R_5$};
\node[] at (axis cs: 1,10^6.5) {\LARGE Cu$\square$};
\node[] at (axis cs: 10^8,10^5.5) {\Huge (a)};
\end{axis}\end{tikzpicture}
\begin{tikzpicture}
\begin{axis}[
	width = 0.7\linewidth, height = 0.45\linewidth,
	%xlabel=\large Photon Number $n$,
    xmode=log,
        xmax = 1*10^9,xmin = 3*10^-2,xtick={1,10^2,10^4,10^6,10^8},ymax = 10^7,
	ylabel=$Q_i$,ymode = log,
	axis x line = bottom,
	minor x tick style={draw=none},
	mark options={mark size=3.5pt},% Ben demands more distinguishable markers
    xticklabels={,,,,},% hide x tick label for this figure
	y tick label style={font=\large}, label style={font=\Large},% Ben demands larger size
        major tick style={very thick, major tick length=5pt, black}, % Ben demands bolder and more visible tickets
        tick align=inside,
	execute at begin axis={\draw (rel axis cs:0,1) -- (rel axis cs:1,1);},
]
\addplot[color = blue, mark = *] table [x=<n>,y=Qi]{AlNoHole_R1.txt};
%\addlegendentry{$R_1$};
\addplot[color = purple, mark = square*] table [x=<n>,y=Qi]{AlNoHole_R2.txt};
%\addlegendentry{$R_2$};
\addplot[color = orange, mark = triangle*] table [x=<n>,y=Qi]{AlNoHole_R3.txt};
%\addlegendentry{$R_3$};
\addplot[color = olive, mark = pentagon*] table [x=<n>,y=Qi]{AlNoHole_R4.txt};
%\addlegendentry{$R_4$};
\addplot[color = teal, mark = diamond*] table [x=<n>,y=Qi]{AlNoHole_R5.txt};
%\addlegendentry{$R_5$};
\node[] at (axis cs: 1,10^6.6) {\LARGE Al$\blacksquare$};
\node[] at (axis cs: 10^8,10^5.8) {\Huge (b)};
\end{axis}
\end{tikzpicture}
\begin{tikzpicture}
\begin{axis}[
	width = 0.7\linewidth, height = 0.45\linewidth,
	xlabel=\large Photon Number $n$,xmode=log,
        xmax = 1*10^9,xmin = 3*10^-2,xtick={1,10^2,10^4,10^6,10^8},ymax = 10^7,
	ylabel=$Q_i$,ymode = log,
	axis x line = bottom,
	minor x tick style={draw=none},
	mark options={mark size=3.5pt},% Ben demands more distinguishable markers
	tick label style={font=\large}, label style={font=\Large},% Ben demands larger size
        major tick style={very thick, major tick length=5pt, black}, % Ben demands bolder and more visible tickets
        tick align=inside,
	execute at begin axis={\draw (rel axis cs:0,1) -- (rel axis cs:1,1);},
]
\addplot[color = blue, mark = *] table [x=<n>,y=Q_i]{AlHole_R1.txt};
%\addlegendentry{$R_1$};
\addplot[color = purple, mark = square*] table [x=<n>,y=Q_i]{AlHole_R2.txt};
%\addlegendentry{$R_2$};
\addplot[color = orange, mark = triangle*] table [x=<n>,y=Q_i]{AlHole_R3.txt};
%\addlegendentry{$R_3$};
\addplot[color = olive, mark = pentagon*] table [x=<n>,y=Q_i]{AlHole_R4.txt};
%\addlegendentry{$R_4$};
\addplot[color = teal, mark = diamond*] table [x=<n>,y=Q_i]{AlHole_R5.txt};
%\addlegendentry{$R_5$};
\node[] at (axis cs: 1,10^6.5) {\LARGE Al$\square$};
\node[] at (axis cs: 10^8,10^5.5) {\Huge (c)};
\end{axis}
\end{tikzpicture}\par}
\caption{Log-log plot of the fitted resonator internal quality factors versus stored photon number for the resonator chip in (a) $\text{Cu}\square$, (b) $\text{Al}\blacksquare$, and (c) $\text{Al}\square$  packages.}
\label{fig:package_all5}
\end{figure}
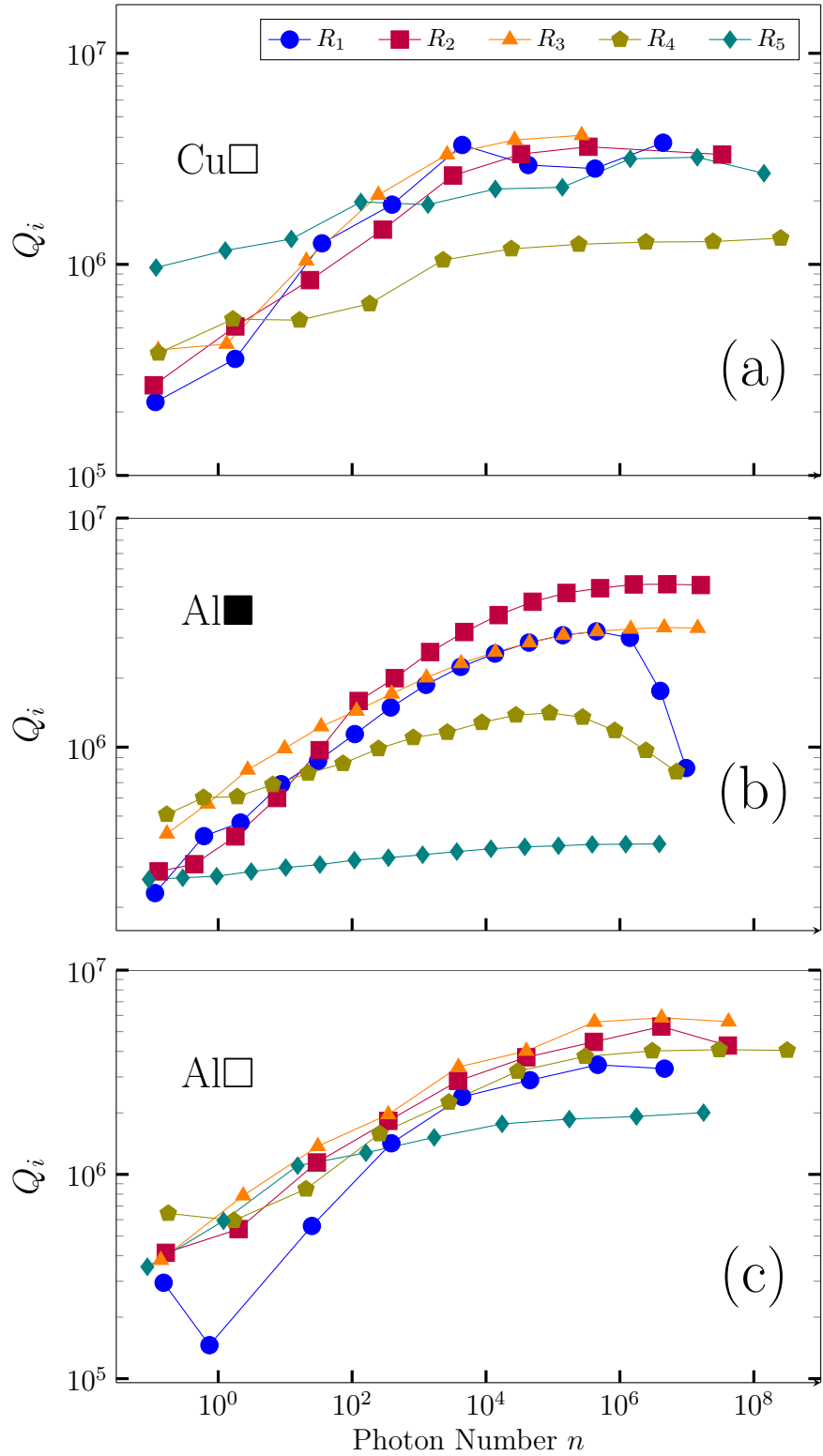

\begin{table}[htbp]
{\centering
\caption{VNA Measurement Settings, the terms with arrow ``$a\rightarrow b$'' means value $a$ was used at the lowest probe power and increased to value $b$.}\label{tab:vna_setting}
\begin{tabular}{|l|l|l|l|l|}
\hline
                                                                                      & $\text{Cu}\blacksquare$ & $\text{Cu}\square$   & $\text{Al}\blacksquare$ & $\text{Al}\square$   \\ \hline
\begin{tabular}[c]{@{}l@{}}Minimum Probe \\ Power (dBm)\end{tabular}                  & -80                     & -80                  & -75                     & -80                  \\ \hline
Power Step                                                                            & 10                      & 10                   & 5                       & 10                   \\ \hline
\begin{tabular}[c]{@{}l@{}}Number of Repetition \\ at Each Probe Power\end{tabular}   & 100                     & 100                  & 50 $\rightarrow$ 5      & 23                   \\ \hline
\begin{tabular}[c]{@{}l@{}}Intermediate Frequency \\ (IF) Bandwidth (Hz)\end{tabular} & 50 $\rightarrow$ 100    & 10 $\rightarrow$ 100 & 100                     & 10 $\rightarrow$ 100 \\ \hline
\end{tabular}\par}
\end{table}
The fitted $Q_i$ versus photon number for the device's first cooldown in $\text{Cu}\blacksquare$ package is shown in the main text. The fitted values from the other packages are shown in Fig.\ref{fig:package_all5}. Note that at the largest applied power for R1 shown in Fig.\ref{fig:package_all5}(b), $Q_i$ begins to decrease as a result of the beginning of nonlinear effects[3].

\section{COMSOL Simulation}\label{sec:comsol}
\begin{figure}[htbp]
	\centering
	\includegraphics[width=0.88\linewidth]{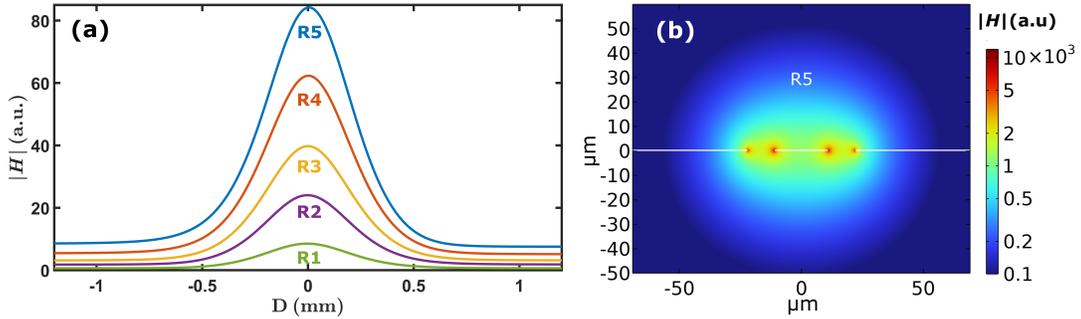}
	\caption{2D COMSOL simulation of the $|H|$ field associated with a TEM wave moving through a CPW. (a)  The magnitude of the $|H|$ field 430 $\mu\text{m}$ below the CPW at the interface of the substrate and the metal package. (b) Cross-sectional view of the $|H|$ field of the TEM mode at the CPW metal surface associated with R5.}\label{fig:comsol_field}
\end{figure}

In addition to the HFSS simulation discussed in the main text, COMSOL was used to simulate the 2D EM fields associated with a TEM wave moving perpendicular to a cross-section of a coplanar waveguide (CPW) (Fig. \ref{fig:comsol_field}(b)). For this simulation, all of the conductors were assumed to be perfect electric conductors. This 2D method of simulating the resonator's EM fields neglects the 3D profile of the device, which for some of our results is important, but provided an intuition of the induced currents underneath the substrate and allowed us to simulate thinner structures, such as thin dielectrics[1] which could contribute to surface loss. 

Fig. \ref{fig:comsol_field}(a) displays the magnitude of the $\left|H\right|$ field at the interface between the bottom of the substrate and the metal package without hole, for R1 to R5 respectively. The $|H|$ field for each resonator has a maximum value, which increases with an increase in the resonator size,  directly under each waveguide. The full width half maximum (FWHM) is independent of resonator size and is approximately $490\ \mu m$, a value that is similar to the substrate thickness.  The magnetic field geometric factors $\gamma$ extracted from COMSOL were within a factor of two of the values calculated from the HFSS simulations.

We  used this same 2D geometry in COMSOL to estimate the effect of surface loss in the main text. For this we assumed three interfaces participating in the loss, metal-substrate (MS), substrate-air (SA), and metal-air (MA) [2] so that the total surface loss is given by
\begin{equation}
    \frac{1}{Q_{SL,i}} = \frac{1}{Q_{MS}} + \frac{1}{Q_{SA}} + \frac{1}{Q_{MA}}.
\end{equation}
For each surface, we assumed a uniform thickness $t$ and calculated the ratio of the energy of the electric field in that layer to the total electric field energy
\begin{equation}
    \frac{1}{Q_k} = \tan{\delta_k}\frac{\int\int \epsilon_k|E|^2}{\sum_j \int\int \epsilon_j|E|^2}, k\in\{\text{MS,SA,MA\}}. \label{eq:sup:comsol_loss}
\end{equation}
For relatively thin  layers, we found that the overall loss in Eq.\ref{eq:sup:comsol_loss} is largely proportional to the term $\delta_k\times t / \epsilon_k$, but independent of its individual components. We used values from Ref.[2] for the loss tangent $\delta_k$ and relative dielectric constant $\epsilon_k$ for each of the three surfaces, performed a few simulations and found $t\sim 0.175\text{nm}$ would give a ``surface loss'' that  approximates the trend observed in the smallest resonators for the Al$\square$ package.

\section{HFSS Simulation}

The HFSS simulation involved a sapphire chip with dimensions $5\times 5\times 0.43\ \text{mm}^3$ and dielectric constant $\epsilon_r = 10.8$. Surrounding the chip with a $0.1\text{mm}$ gap on all sides for the two Cu packages  was a PCB with outer dimensions $10\times 8 \text{mm}^2$. The top and bottom of the PCB were defined as a perfect electric conductor with a thickness of $17\ \mu\text{m}$. A series of vertical vias, set as perfect conductors, was used to connect the top and bottom layers. In between the two PCB metal layers was a 396 $\mu\text{m}$ thick dielectric layer with  $\epsilon_r = 6$ and a loss tangent $\delta = 0.0026$ as specified by the manufacturer at room temperature. Our simulations suggest that the PCB dielectric  negligibly contributed to the loss so  its contribution was omitted in the main text. While Fig.1 in the main text displays all 5 resonators, their geometric factors $\gamma$ were individually simulated with only one resonator present to save simulation resources.

In simulations of $\text{Cu}\square$ and the two Al packages, we found that the presence of the glue reduces $\gamma_\text{Base}$ and $\gamma_\text{PCB}$ due to the glue partially covering those surfaces. To account for this, we performed two simulations, one with glue present and one absent. The extracted geometric factors are $\gamma_\text{Base, glue absent}$,$ \gamma_\text{PCB, glue absent}$, $\gamma_\text{Base, glue present}$, $\gamma_\text{PCB, glue present}$, and $\gamma_\text{glue}$.
The contributions in Fig.4 in the main text were calculated as
\begin{align}
Q_\text{Base}^{-1} =& \frac{R_{S,\text{base}}}{\omega \mu_0}\times \gamma_\text{Base, glue absent}\\
Q_\text{PCB}^{-1} =& \frac{R_{S,\text{PCB}}}{\omega \mu_0}\times \gamma_\text{PCB, glue absent}\\
Q_\text{Glue}^{-1} =& \frac{R_{S,\text{base}}}{\omega \mu_0}\times \gamma_\text{Base, glue present}+ \frac{R_{S,\text{PCB}}}{\omega \mu_0}\times \gamma_\text{PCB, glue present}\notag\\&+ \frac{R_{S,\text{glue}}}{\omega \mu_0}\times \gamma_\text{glue}-Q_\text{Base}^{-1}-Q_\text{PCB}^{-1}.
\end{align}

For $\text{Cu}\blacksquare$ package, to prevent overestimating loss from the glue, we assumed the glue covered 70\% of the bottom side of the substrate, with its contribution in Fig.4 in the main text calculated from 
\begin{equation} Q_\text{Glue}^{-1} = 70\% \times \frac{R_{S,\text{glue}} - R_{S,\text{base}}}{\omega \mu_0}\times \gamma_\text{Base}.\end{equation}

The lid of $\text{Cu}\blacksquare$ package was also made of normal metal copper. It had a geometric factor $\gamma_\text{Lid}$ two orders of magnitude smaller than that of package base $\gamma_\text{Base}$ so we omitted it in the calculations.

%\nocite{*}
%\bibliography{aip}% Produces the bibliography via BibTeX.
%\printbibliography